\documentclass[prd,aps,a4paper,superscriptaddress,twocolumn,nofootinbib]{revtex4} %

\usepackage{graphicx}
\usepackage{color}
\usepackage{dcolumn}
\usepackage{bm}
\usepackage{slashed}
\usepackage{amsmath}
\usepackage{latexsym}
\usepackage{amssymb}
\usepackage{mathrsfs}
\usepackage{amsfonts}
\usepackage{url}
\allowdisplaybreaks
\usepackage{diagbox}
\usepackage{multirow}
\usepackage{makecell}
\usepackage{natbib}
\usepackage{xcolor}
\usepackage{comment}
\usepackage[colorlinks=true,linkcolor=blue,urlcolor=blue,citecolor=blue]{hyperref}

\begin{document}
\title{Extraction of binary neutron star gravitational wave waveforms from Einstein Telescope using deep learning}
	
\author{CunLiang Ma}
\affiliation{School of Information Engineering, Jiangxi University of Science and Technology, Ganzhou, 341000, China}
\affiliation{Jiangxi Provincial Key Laboratory of Multidimensional Intelligent Perception and Control, Ganzhou, 341000, China}

\author{XinYao Yu}
\affiliation{School of Information Engineering, Jiangxi University of Science and Technology, Ganzhou, 341000, China}
		
\author{Zhoujian Cao
\footnote{corresponding author}} \email[Zhoujian Cao: ]{zjcao@amt.ac.cn}
\affiliation{Institute of Applied Mathematics, Academy of Mathematics and Systems Science, Chinese Academy of Sciences, Beijing 100190, China}
\affiliation{School of Fundamental Physics and Mathematical Sciences, Hangzhou Institute for Advanced Study, UCAS, Hangzhou 310024, China}

\author{Mingzhen Jia}
\affiliation{School of Information Engineering, Jiangxi University of Science and Technology, Ganzhou, 341000, China}
\affiliation{Jiangxi Provincial Key Laboratory of Multidimensional Intelligent Perception and Control, Ganzhou, 341000, China}

\begin{abstract}
        In the future, the third generation (3G) gravitational wave (GW) detectors, exemplified by the Einstein Telescope (ET), will be operational. The detection rate of GW from binary neutron star (BNS) is expected to reach approximately $10^4$ per year. To address the challenges posed by BNS GW data processing for 3G GW detectors, this paper explores the extraction of BNS waveforms from ET. Drawing inspiration from SPIIR's matched filtering approach, we introduce a novel framework leveraging deep learning for BNS waveform extraction. By integrating denoised outputs of time-delayed strain, we can reconstruct the embedded BNS waveform. We have established three distinct BNS GW denoising models, each tailored to address the early inspiral, later inspiral, and merger phases of BNS GW, respectively. To further regulate the waveform shape, we propose the Amplitude Regularity Model that takes denoised output as input and regulated waveform as output. The experiments conducted on test data demonstrate the efficacy of the denoising models, the Amplitude Regularity Models, as well as the overall waveform construction method. To the best of our knowledge, this marks the first instance of deep learning being utilized for the task of BNS waveform extraction. We believe that the proposed method holds promise for early warning, searching, and localization of BNS GWs.
\end{abstract}

\maketitle

\section{Introduction\label{section:1}}

After conducting the initial three observation runs, the International Gravitational-wave Network (IGWN) \cite{1} has accurately detected more than 90 confident gravitational wave (GW) events \cite{2,3,4,5}. Among the confidently detected GW events, GW170817 and GW190425 specifically stem from the merging of binary neutron stars (BNSs). Meanwhile, other confidently identified events originate from either binary black hole (BBH) or neutron star-black hole (NSBH) mergers.

The observation of GW170817 was witnessed both in GW and electromagnetic spectra \cite{6,7}. This groundbreaking observation ushered in the dawn of Multi-Messenger Astrophysics that leverages observations across electromagnetic radiation, gravitational waves, cosmic rays, and neutrinos. GW170817 furnished conclusive proof that BNSs are one of the driving force behind short gamma ray bursts, marking the first direct evidence of this phenomenon, and it definitively established that GWs propagate at speeds virtually indistinguishable from that of light \cite{8}. Most importantly, the discovery of GW170817 has played a pivotal role in illustrating that gravitational wave observations have the potential to deduce the tidal deformability of neutron stars \cite{9,10,11}.

Conventional GW search methods primarily rely on a technique called template matched filtering \cite{12,13,14,15,16,17}. This approach typically employs a template bank with extensive waveforms, each distinguished by various compact binary parameters, including but not limited to component masses and/or spins. Specifically, GW searches employing matched filtering techniques are currently focused on a distinct segment of the available parameter space. As mentioned in the literature \cite{18}, the main emphasis is on a 4D parameter space, which represents compact binary sources where the spin-aligned components orbit in quasi-circular paths. 

In the future, third-generation (3G) detectors, such as Einstein Telescope (ET) \cite{19} and Cosmic Explorer (CE) \cite{20}, will work. These detectors promise remarkable sensitivity advancements \cite{21}, elevating it by an order of magnitude. Furthermore, they are expected to considerably widen the bandwidth, reaching both lower and higher frequencies. In these cases, GW detectors have the potential to explore a broader 9D parameter space. In such a scenario, the computational demands of these low-latency GW searches will be exceedingly high, potentially posing significant challenges. 

To address the computational efficiency challenge, numerous studies have concentrated on leveraging deep learning for GW searches \cite{22,23,24,25,26,27,28,29,30,31,32,33,34,35,36,37}. In the case of the BNS merger, search efficiency holds utmost importance due to the critical need for swift follow-up observations to successfully detect their electromagnetic counterparts. Several studies have shown that deep learning can be efficiently used to search for BNS mergers \cite{38,39,40,41,42,43,44}, specifically in detecting pre-merger alerts from GWs emitted by BNSs \cite{45,46,47}. The application of deep learning in this context has been thoroughly explored, demonstrating its potential in identifying such events.

All BNS GW searches based on deep learning employ an end-to-end classification approach. Alternatively, the matched filtering method for BNS GW searches has the capability to exhibit the fluctuation of the signal-to-noise ratio (SNR) over time. Furthermore, the output generated through matched filtering can facilitate subsequent endeavors, such as source localization \cite{48}. On the other hand, GW search methods based on end-to-end classification often face difficulties in providing the prior information required for tasks such as wave source localization and parameter estimation. The traditional matched filtering for BNS GW detection also faces challenges. Even a minor alteration in BNS masses can result in a significant waveform mismatch in this area, primarily due to the increased number of cycles within the frequency band that require precise alignment.

Fortunately, in 2023, a new task for envelope extraction was proposed, capable of predicting the hidden gravitational wave signal's envelope \cite{49}. This task was employed to further validate search results by comparing the coalescence times across various detectors. Typically, most deep learning-based detection methods treat all time segments, regardless of whether they contain gravitational wave information, equally. However, recently, we leveraged the envelope extraction network to identify significant data segments. Subsequently, we utilized these key segments to predict templates using denoising \cite{50}. These predicted templates then facilitated matched filtering. The denoising of BBH case has undergone extensive research \cite{51,52,53,54}. Therefore, the framework can be conveniently utilized for BBH search. Currently, there is no existing research on denoising techniques specifically designed for BNSs. As a result, the application of the MSNRnet framework for BNS searching remains challenging.

If the denoising process can accurately predict the BNS GW template, the waveform mismatch issue inherent in matched filtering will be effectively resolved. Template prediction is just one among several applications of BNS GW denoising. The GW waveforms emitted by the BNS system carry vital information about the internal structure of neutron stars, reflected through the tidal deformability parameter, which is dependent on the equation of state of the neutron star. Additionally, the denoised BNS GW output aids in exploring the deformability of neutron stars. 

In this study, we concentrate primarily on addressing the denoising challenge posed by binary neutron stars. It's crucial to note that denoising techniques developed for BBH cannot be seamlessly applied to BNS. This is primarily due to distinct characteristics inherent to each type of binary system. For instance, BBH signals typically last less than 2 seconds within the detector's sensitivity range, whereas BNS signals can persist for over 100 seconds. Additionally, the GW emitted during a BNS merger have a higher peak frequency compared to those from a BBH merger. Furthermore, given the same matched filtering SNR, the peak amplitude of GW from a BNS is notably weaker than that of a BBH.

For the greater challenge of GW data processing in the future 3G detectors, in this work we focus on the BNS waveform construction based on 3G detectors taking ET as an example. The ET will enable us to embark on an exploration of the Universe through GWs, tracing its cosmic history all the way back to the cosmological dark ages. This groundbreaking endeavor promises to illuminate unresolved issues in fundamental physics and cosmology. The ET is expected to have a detection rate of approximately $10^4$ to $10^6$ events per year for BBHs and $7 \times 10^4$ events per year for BNSs \cite{55,56}. Recently, the detection of GW data from ET using deep learning methods has garnered significant attention, including BBH \cite{57,58}  and Cosmic String Cusps \cite{59}.

For the long time-duration and high peak frequency of the BNS signal, in terms of computational complexity, it is virtually impossible to denoise all information from a deep learning model directly. Motivated by the template bank construction methods of traditional matched filtering pipelines SPIIR, we propose a new framework for BNS waveform construction. The SPIIR construct the template via summation of time shifted infinite impulse response (IIR) filters. In our framework we construct the waveform by combine of time shifted denoising results by different deep filters. Three sub-models (deep filters) are used for the overall waveform construction task. The first model pertains to the early inspiral phase of the signal, the second model focuses on the later inspiral stage of the signal, while the third model focuses on near merger stage that encompasses the inspiral leading up to the merger, the merger itself, and the subsequent ringdown stages of the BNS signal. We believe that the denoising model specifically designed for the early inspiral phase can serve as an early warning system for BNS events. Additionally, the comprehensively constructed waveform can be effectively utilized for template creation.

The sampling frequencies of our different denoising models are tailored to the specific frequency ranges of each stage of the signal. The overall signal may cover a wide frequency spectrum, but unique stages such as early inspiral, late inspiral, and merger display distinct frequency characteristics. Therefore, we utilize different sampling frequencies for the corresponding denoising models to address the specific needs of each stage.

\section{DATA FOR TRAINING AND TESTING\label{section:2}}

\begin{figure}[ht]
        \includegraphics[scale=0.3]{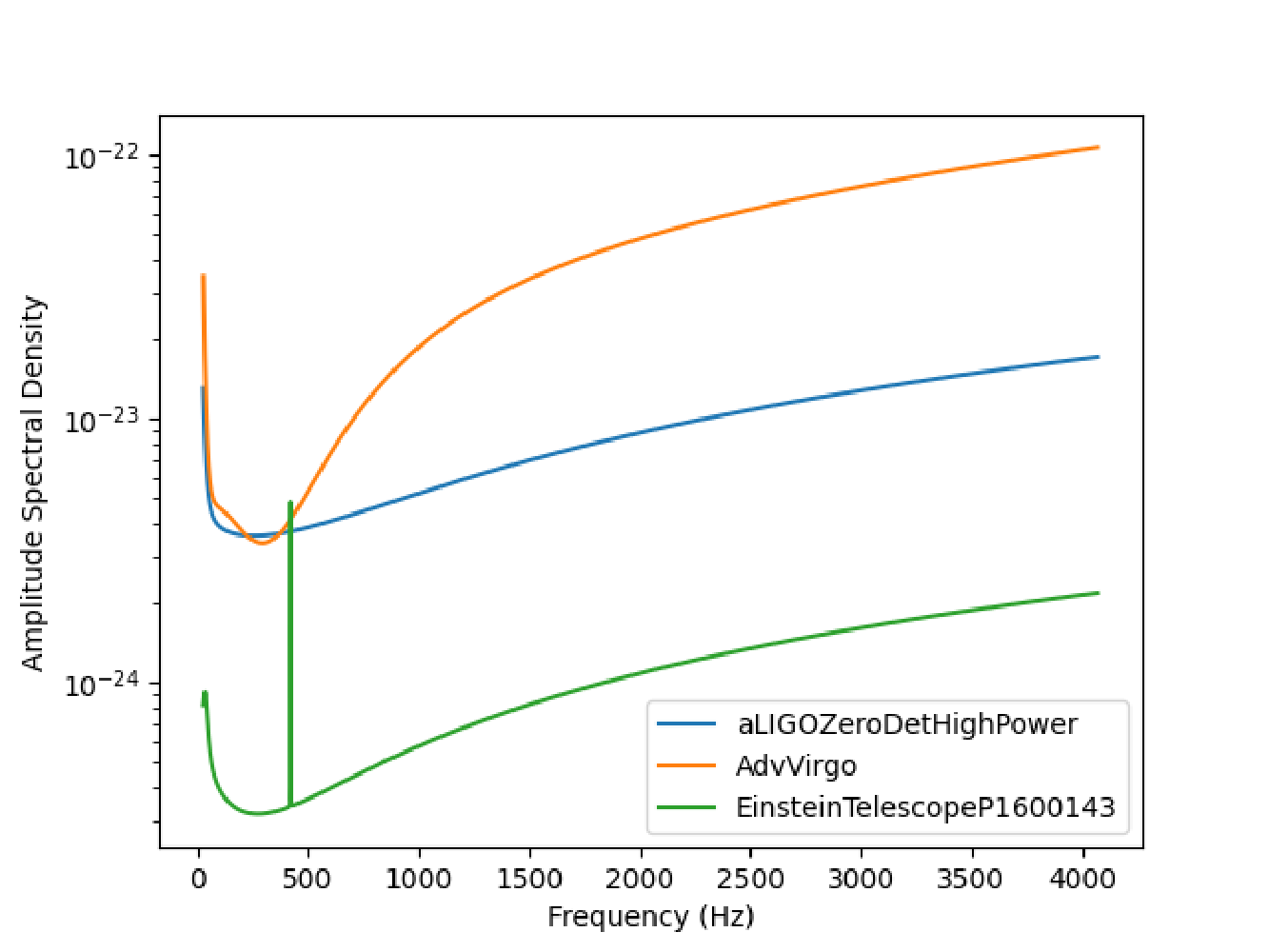}
        \caption{\label{fig:1} The ASD curves of Einstein Telescope, Advanced LIGO and Advanced Viego.}
\end{figure}

The Einstein Telescope will be composed of three detectors positioned in a triangular layout, thereby enabling the simultaneous collection of three detectors strain data streams. In this work we only focus on one detector. We utilize the PyCBC package \cite{60,61,62,63,64} to synthesize data for training, validation, and testing purposes. The strain s(t) obtained by the detector can be modeled as

\begin{equation}
        s(t)=h(t)+n(t),\label{eq:1}
\end{equation}
where $h(t)$ denotes the GW signal and $n(t)$ denotes the background noise. $h(t)$ can be obtained by a linear combination of $h_+(t)$ and $h_\times(t)$ that can be simulated using physical models. In this work, we use IMRPhenomD\_NRTidal to generate $h_+(t)$ and $h_\times(t)$. For the simulation, we established a distance of 100 Mpc between the Earth and the BNS source. During the training phase, we adjusted the amplitude of $h(t)$  to attain a randomly sampled target SNR falling within the range of 20 to 40. The masses of the two neutron stars in the BNS system were randomly selected from the interval of  $(1M_\odot, 2M_\odot)$ , while their dimensionless spins were sampled randomly between 0 and 0.998. The right ascension and declination angles were chosen from a uniform distribution across a sphere. We cut each signal to 100 seconds duration. The signals are sampled at a rate of 8192Hz. The noise was produced through the application of the power spectrum density (PSD) associated with the Einstein Telescope. This PSD serves as an indicator of the detector's design sensitivity. Specifically, we relied on the PSD designated as EinsteinTelescopeP1600143 to create the noise. Fig. \ref{fig:1} illustrates the variation of amplitude spectrum density (ASD, which is the square root of PSD) with respect to frequency. For comparative purposes, the ASD curves for advanced LIGO (specifically, aLIGOZeroDetHighPower) and advanced Virgo (AdvVirgo) are also displayed alongside.

We organize our data using data packages, where each package contains one noise sample lasting 256 seconds and ten signal samples, each lasting 100 seconds. In total, we have generated 30,000 such packages, with 24,000 designated for training, 3,000 for validation, and the remaining 3,000 for testing. During the training stage, every sample is produced through instant synthesis by randomly selecting a signal, sliding it randomly, adjusting the SNR randomly, and finally, combining it with a randomly chosen noise slice. Consequently, each package has the capability to generate thousands of distinct samples for training purposes.

\section{DEEP LEARNING FRAMEWORK FOR THE BNS WAVEFORM CONSTRUCTION\label{section:3}}

\begin{figure}[htbp]
        \includegraphics[scale=0.2]{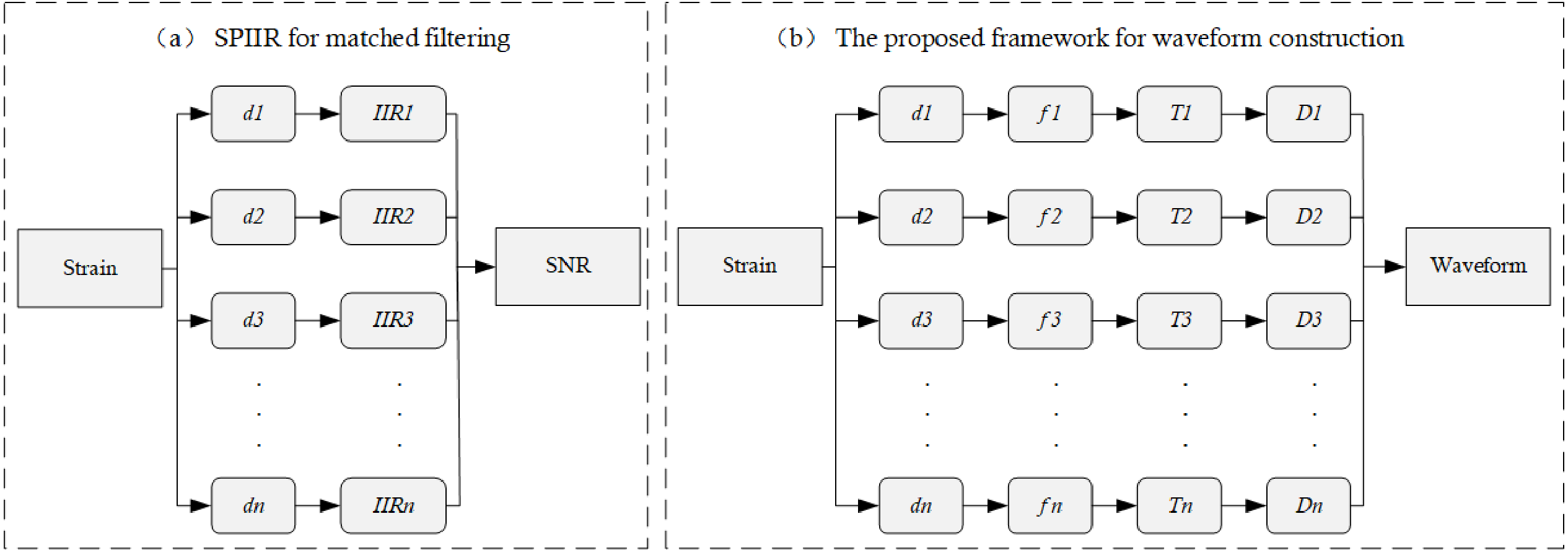}
        \caption{\label{fig:2}A schematic of the SPIIR method and the proposed method. $di$ denotes the time delay of the ith channel, $fi$ signifies the cut-off frequency of the low-pass filter, $Ti$ represents the sampling period, and $Di$ designates the ith denoising model.}
\end{figure}

In this section, the proposed method for BNS waveform construction will be systematically introduced. We segment the entire time scale into numerous smaller channels, enabling each denoising model to concentrate solely on the response within its designated channel. This strategy is inspired by the SPIIR matched filtering pipeline. The diagram illustrating SPIIR for matched filtering and our proposed denoising framework (waveform construction) is presented in Fig. \ref{fig:2}. In SPIIR, the waveform is estimated by adding up time-shifted, exponentially increasing sinusoids, which are formed using single-pole IIR filters. In contrast, our method estimates the waveform by summing the denoised outputs of time-shifted segments. Initially, for both SPIIR and our approach, the input signal is divided into multiple channels. Each channel $i$ undergoes a specific time delay of $di$. In SPIIR, the time-delayed channel $i$ subsequently passes through the respective IIR filter (IIR$i$). The IIR filters’ outputs are then aggregated to determine the SNR output.

\begin{figure}[htbp]
        \includegraphics[scale=0.2]{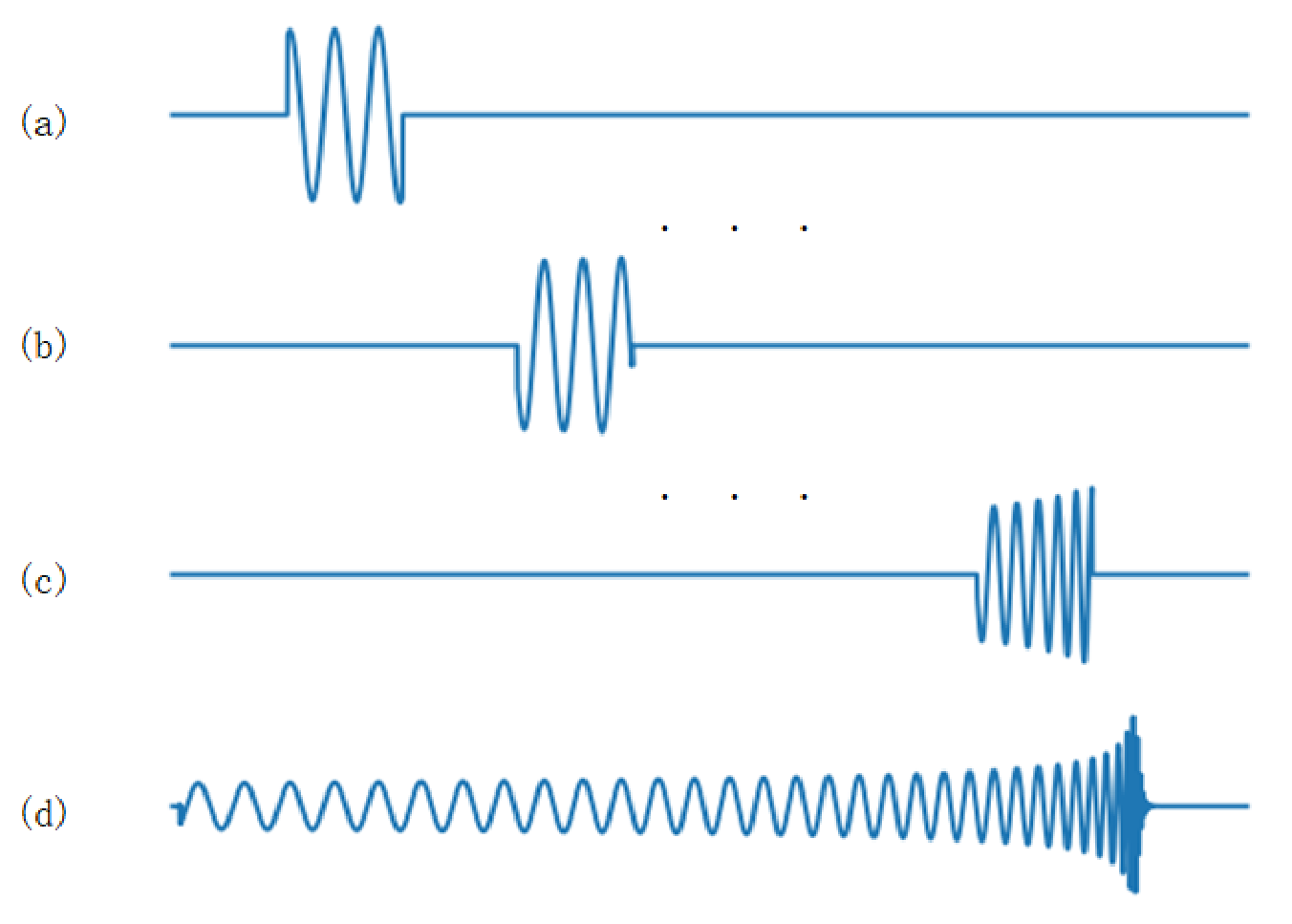}
        \caption{\label{fig:3}An illustrative diagram presents the capability of constructing the final waveform using denoised outputs. The top three panels, labeled (a), (b), and (c), exhibit three instances of denoised outputs. Panel (d) displays the resulting constructed waveform. Please bear in mind that this figure serves solely as a visual aid for explanation.}
\end{figure}

In our method, the $i$-th time-delayed channel is processed through a Butterworth low-pass filter with a cutoff frequency of $\mathit{fi}$. Subsequently, the filtered strain from the $i$-th channel is resampled using a sampling period of $\mathit{Ti}$, where the relationship between $\mathit{Ti}$ and $\mathit{fi}$ is defined as $2\mathit{fi}=1/\mathit{Ti}$. The resampled strain is then fed into the corresponding denoising process. Finally, the waveform is reconstructed from the outputs of the denoising process. Fig. \ref{fig:3} illustrates the process of combining denoised outputs to construct the final waveform. In this article, we utilize three channels as an illustrative example to explore the process of constructing BNS waveforms. The first channel is dedicated to the early inspiral stage, the second channel centers on the later inspiral stage, while the third channel concentrates on the near merger stage. These three stages correspond to three unique denoising processes, specifically Process I, Process II, and Process III. Each process consists of a denoising model and an amplitude regularity model. Suppose the input of the denoising process of channel $i$ is $s^i=n^i+h^i$, where $h^i \in R^{40960}$ denotes the buried signal. The denoising process can be formulated as follows

\begin{equation}
        \hat{h}_D^i=Denoise_i(s^i|W_{D,i}),\label{eq:2}
\end{equation}
\begin{equation}
        \hat{h}^i=Amplitude_i(\hat{h}_D^i|W_{A,i}),\label{eq:3}
\end{equation}
where $W_{D,i}$ and $W_{A,i}$ are the trainable parameters of the corresponding denoising model and amplitude regularity model respectively and can be optimized by

\begin{equation}
        W_{D,i}=\underset{W_{D,i}}{\operatorname*{argmin}}MSE(\hat{h}_{D}^{i},h^{i}).\label{eq:4}
\end{equation}
\begin{equation}
        W_{A,i}=\underset{W_{A,i}}{\operatorname*{argmin}}MSE(\widehat{h}^{i},h^{i}).\label{eq:5}
\end{equation}

\begin{table}[htbp]
        \caption{\label{tab:table1}%
        The sampling frequency, time duration of the three denoising models
        }
        \begin{ruledtabular}
        \begin{tabular}{cccc}
        \makecell[c]{Denoising \\ process}&\makecell[c]{Sampling \\ frequency}&\makecell[c]{Time duration}&\makecell[c]{Input/output \\ shape}\\  \hline
        Process I & 512 Hz & 80 s & $R^{40960}$\\
        Process II & 2048 Hz & 20 s & $R^{40960}$\\
        Process III & 8192 Hz & 5 s & $R^{40960}$\\
        \end{tabular}
        \end{ruledtabular}
        \end{table}

The structures of the models of each processes are the same, differences are the sampling frequency and time duration and these information of the three models are shown in Table \ref{tab:table1}. The structure of denoising model, amplitude regularity model will be detailed in the following subsections.

\subsection{Denoising model\label{Denoising model}}

\begin{figure*}[htbp]
\includegraphics[scale=0.35]{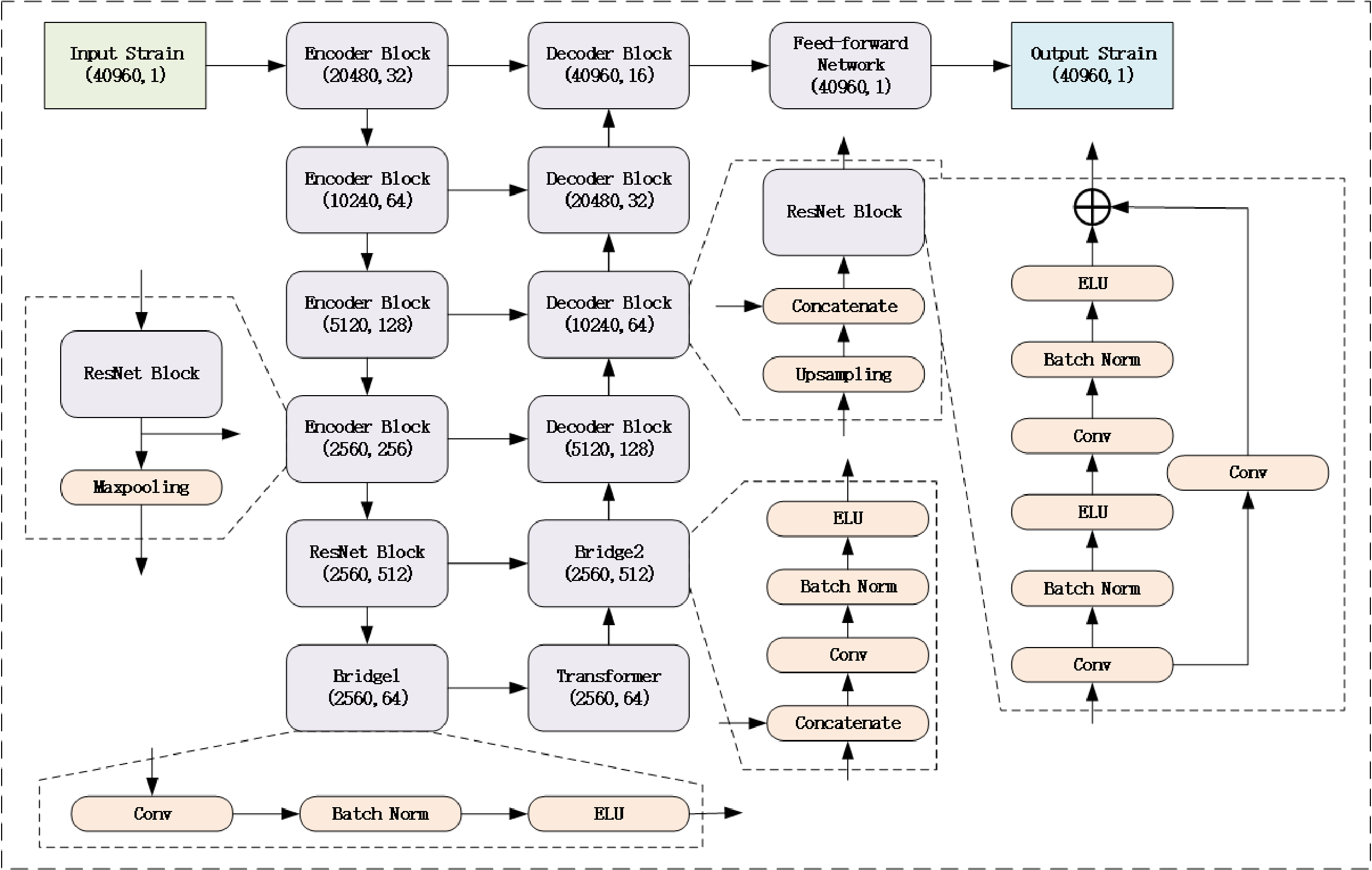}
\caption{\label{fig:4}The network structure of the Denoising model in the current work. The wide boxes refer to individual operations, while the narrow boxes represent sets of multiple operations. The parentheses below the boxes indicate the output dimensions of the operation group, representing the dimension of a single data point and the number of channels, respectively. The XOR symbol represents a residual structure.}
\end{figure*}

The structure of the denoising model is illustrated in Fig. \ref{fig:4}, and it consists of an Encoder-Decoder Module and a Feed-forward Network. The Encoder Module is made up of 4 Encoder blocks, 1 ResNet block, and 1 Bridge block, while the Decoder Module comprises 4 Decoder blocks, 1 Bridge block, and 1 Transformer block. In the subsequent sections, we will delve into the specifics of the Encoder, Bridge, Transformer, and Decoder blocks.

\subsubsection{Encoder block\label{Encoder block}}

The role of the Encoder is to extract features from the input data in the network. This is achieved through a ResNet block, followed by a Max-pooling layer($Maxpooling$). The ResNet block consists of two Convolutional layers and a Residual connection. Each Convolutional layer comprises of a Convolutional($Conv$), a Batch Normalization($BatchNorm$), and an Exponential Linear Unit($ELU$) activation function. The structure of the ResNet block can be formalized as

\begin{equation}
        X_i'=ELU(BatchNorm(Conv_{(r_{in},r_{out})}(X_{i-1}))),\label{eq:6}
\end{equation}
\begin{equation}
        X_i''=ELU(BatchNorm(Conv_{(r_{out},r_{out})}(X_i'))),\label{eq:7}   
\end{equation}
\begin{equation}
        X_{i}=Maxpooling(X_{i}''\oplus Conv_{(r_{in},r_{out})}(X_{i-1})),\label{eq:8}  
\end{equation}
Where $X_i$ represents the output of the $i$-th Encoder block. $X_i'$ and $X_i''$ represent the intermediate features of each ResNet block. $r_{in}$ and $r_{out}$ denote the number of input channels and output channels, respectively. We stipulate that the sizes of convolutional kernels, filters, strides and maxpooling are specified as follows:$\{(64,64,32,32,16), (32,64,128,256,512), 1, 2\}$.

\subsubsection{Bridge block\label{Bridge block}}

We have additionally designed Bridge blocks to connect the UNet and Transformer. The purpose of Bridge I is to reduce the number of channels in the features in order to save computational resources. The role of Bridge II is to create skip connection that simultaneously preserves the output features of both the Encoder and Transformer, while also restoring the number of channels. The structure of the Bridge blocks can be formalized as

\begin{equation}
        X_{b1}=ELU(BatchNorm(Conv_{(b1_{in},b1_{out})}(X_{5}))),\label{eq:9}
\end{equation}
\begin{equation}
        X_{b2}'=Conv_{(b2_{in},b2_{out})}Concatenate(X_5,X_t),\label{eq:10}
\end{equation}
\begin{equation}
        X_{b2}=ELU(BatchNorm(X_{b2}')),\label{eq:11}
\end{equation}
Where $X_{b1}$ and $X_{b2}$ respectively represent the output of Bridge I and Bridge II. $X_{b2}'$ represents the intermediate feature of Bridge II block. $X_t$ represents the output of Transformer. Akin to $r_{in}$ and $r_{cout}$, $b1_{in}$, $b1_{cout}$, $b2_{in}$ and $b2_{cout}$ denote the input and output channels of each Bridge block. The skip connection is referred as $Concatenate(\cdot)$. We stipulate that the sizes of convolutional kernel, filters, strides and maxpooling are specified as follows:$\{(1,16), (64,512), 1, 2\}$.

\subsubsection{Transformer block\label{Transformer block}}

\begin{figure}[htbp]
        \includegraphics[scale=0.3]{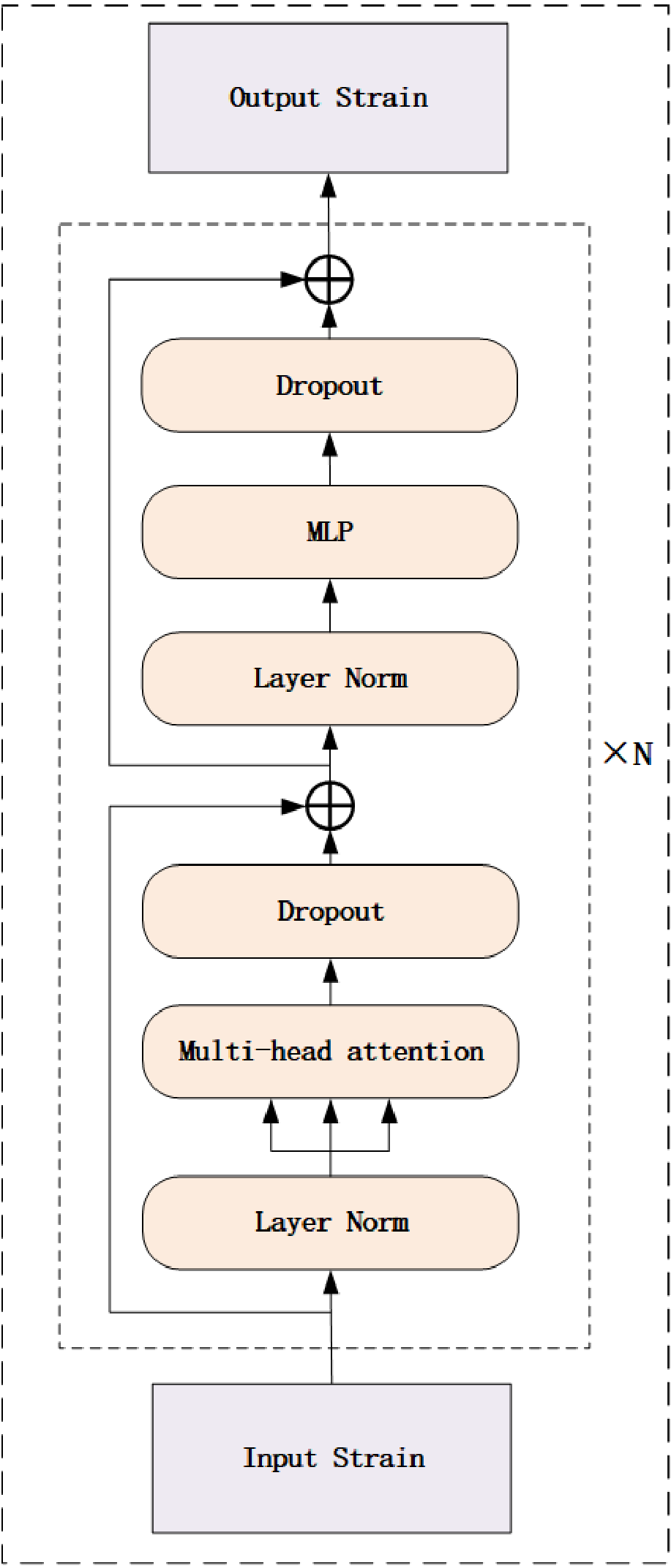}
        \caption{\label{fig:5}The network structure of the transformer. N represents the number of blocks in the Transformer architecture.}
\end{figure}

BNS signal is a long-term sequential signal, and capturing its long-term dependencies is a key focus in denoising research. The self-attention mechanism of the Transformer can effectively capture long-term dependency relationships. The self-attention mechanism computes interaction relationships between time slices, aggregates them as output, and can dynamically aggregate relevant features based on input content, effectively capturing long-range correlations. This is why we introduce the Transformer after the Encoder. The structure of the Transformer we used is shown in Fig. \ref{fig:5}.

To prevent overfitting of the model, we add Dropout in the Transformer. The structure of the Transformer blocks can be formalized as

\begin{equation}
        X_{(t,i)}'=MHA(LayerNorm((X_{(t,i-1)}))),\label{eq:12}
\end{equation}
\begin{equation}
        X_{(t,i)}''=Dropout(X_{(t,i)}')\oplus X_{(t,i-1)},\label{eq:13}
\end{equation}
\begin{equation}
        X_{(t,i)}'''=MLP(LayerNorm((X_{(t,i)}''))),\label{eq:14}
\end{equation}
\begin{equation}
        X_{(t,i)}=Dropout(X_{(t,i)}''')\oplus X_{(t,i)}'',\label{eq:15}
\end{equation}
Where $X_{(t,i)}$ represents the output of $i$-th Transformer block, and $X_{(t,i)}'$, $X_{(t,i)}''$, $X_{(t,i)}'''$ represent intermediate features of the $i$-th Transformer block. Additionally, $LayerNorm(\cdot)$ is the operation of Layer Normalization. $MHA(\cdot)$ refers to Multi-head attention and the details are detailed in \cite{65}. We respectively specify the number of Transformer blocks, Attention Heads, and Dropout rate as 3, 4, and 0.1. For the $MLP$, we follow a two-layer linear transformation structure with a Rectified Linear Unit(ReLU) activation function between the two layers. The units of the linear layer is specified as 128.

\subsubsection{Decoder block\label{Decoder block}}

The Decoder is used to upsample the encoded features and reconstruct them back to the input strain size, providing sufficient signal reconstruction information for the Feed-forward Network. The structure of the Decoder consists of upsampling, skip connections, and residual convolutions. The Decoder first upsamples the output features from the previous layer, then connects it with the output from the same layer of the Encoder, and finally passes through a residual convolution structure identical to that of the Encoder. We stipulate that the sizes of convolutional kernels, filters , strides and upsampling are specified as follows: $\{(32, 32, 64, 64), (128, 64, 32, 16), 1, 2\}$.

\subsubsection{Feed-forward Network\label{Feed-forward Network}}

\begin{figure*}[htbp]
        \includegraphics[scale=0.3]{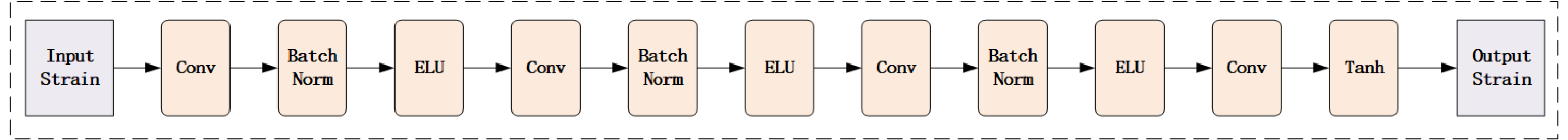}
        \caption{\label{fig:6}The network structure of the Feed-forward Network}
\end{figure*}

The Feed-forward Network obtains signal information from the Decoder and reconstructs the waveform. The structure of the Feed-forward Network we used is shown in Fig. \ref{fig:6}. We stipulate that the sizes of convolutional kernels, filters, and strides are specified as follows: $\{(32, 32, 16, 16), (32, 16, 8, 1), 1\}$.

\subsection{Amplitude Regularity model\label{Amplitude regularity model block}}

\begin{figure*}[htbp]
        \includegraphics[scale=0.4]{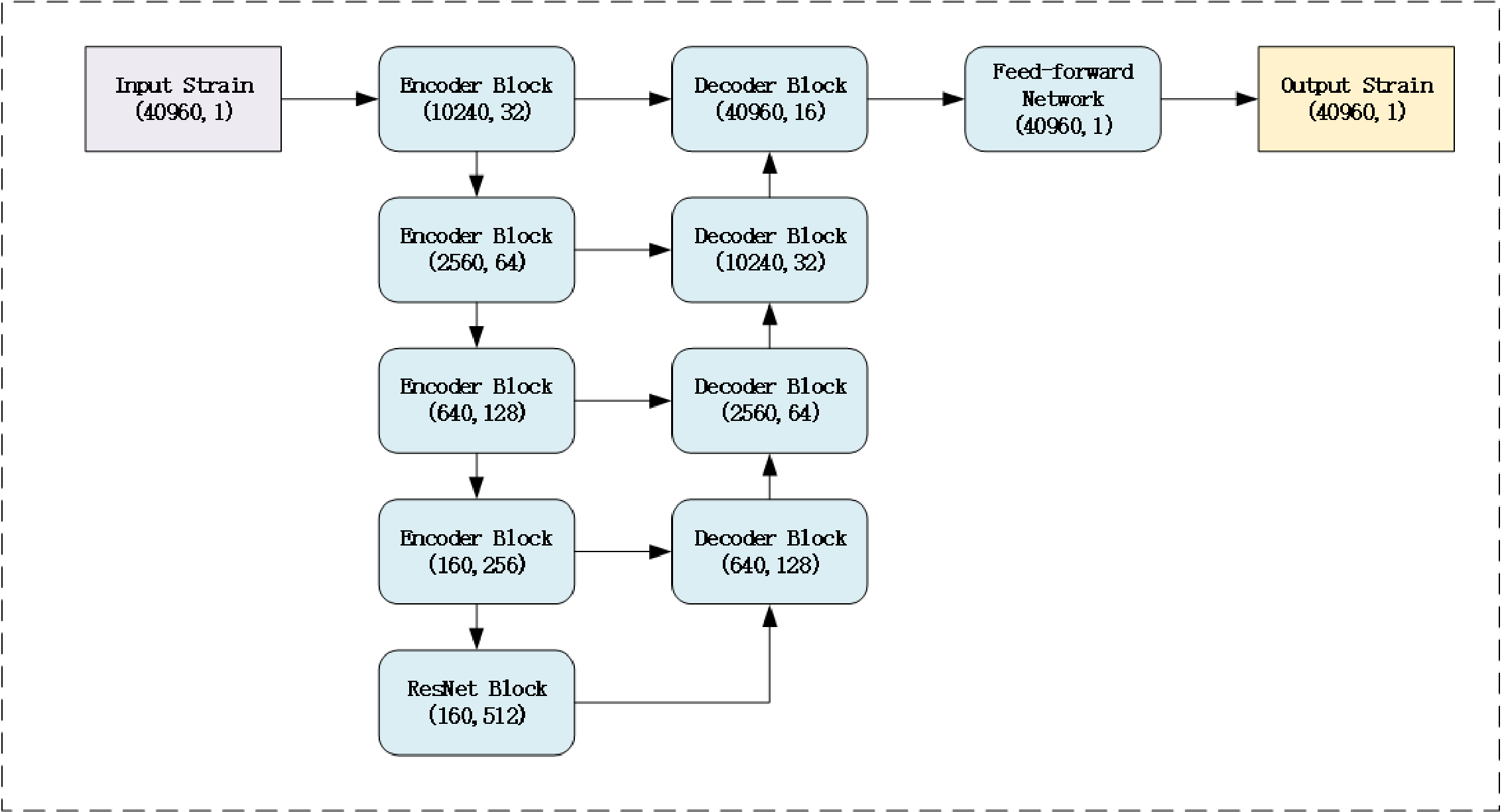}
        \caption{\label{fig:7}The network structure of the Amplitude Regularity model. It is similar to the Denoising model, but removes the Transformer and Bridge structures, and increases the pooling size to 4.}
\end{figure*}

When we test our denoising model in practice, we observe that the denoised results often exhibit amplitude reduction in certain areas. To optimize the local structure of the waveform, we prune the denoising model and construct a new architecture, which we call the Amplitude Regularity Model. This model removes the Transformer and Bridge components and adjusts the pooling and upsampling dimensions to 4. The network structure of Amplitude Regularity Model is shown in FIG. \ref{fig:7}.

\subsection{Waveform construction method\label{Waveform construction method}}

\begin{figure}[htbp]
        \includegraphics[scale=0.5]{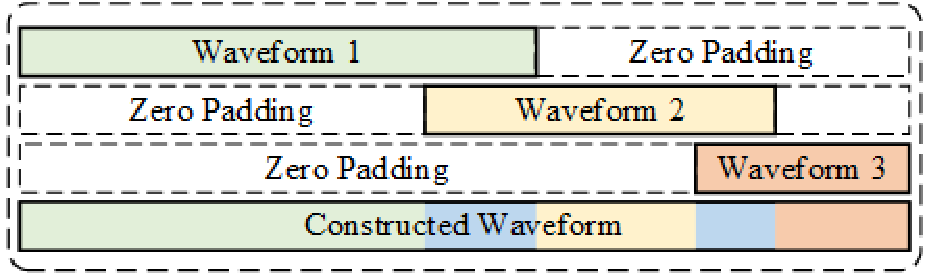}
        \caption{\label{fig:8}The diagram of waveform construction}
\end{figure}

After the previous deep learning processing, we obtain three denoised waveforms (Waveform 1, Waveform 2, and Waveform 3) for different channels with sampling rates of 512 Hz, 2048 Hz, and 8192 Hz, respectively. In this subsection, we detail the waveform construction method. The construction method shown in Fig \ref{fig:8}, takes Waveform 1, Waveform 2, and Waveform 3 as input and Constructed Waveform as output. Due to the varying sampling frequencies of each denoised waveform, Waveform 1 and Waveform 2 must undergo upsampling to 8192 Hz prior to their combination. For the upsampling process, spline interpolation is utilized. Suppose the upsampled Waveform 1, Waveform 2 and Waveform 3 can be denote as $h_1(t)$, $h_2(t)$ and $h_3(t)$. The amplitudes of the three waveforms may not be aligned, thus necessitating amplitude adjustments prior to their combination. Specifically, we must adjust the amplitude of $h_2(t)$ to match that of $h_3(t)$, and similarly, adapt the amplitude of $h_1(t)$ to align with the adapted $h_2(t)$.

Here, we introduce the combination of $h_1(t)$, $h_2(t)$ and $h_3(t)$. We initially add zeroes to the three waveforms to ensure that they all possess the same length. Define the function $I_{ij}(t)$ (refer to Eq. (\ref{eq:16})) which can ascertain whether waveforms $i$ and $j$ exhibit any temporal overlap.  

\begin{align}
        {I_{ij}(t)=\begin{cases}1, \hspace{0.25cm} h_i \ and \ h_j \ at \ time \ t \ are \ overlap\\
                0, \hspace{0.25cm} otherwise\\
        \end{cases}. \label{eq:16}}
\end{align}

% \begin{align}
%         I_{ij}(t)=\left\{\begin{cases}1, h_i and h_j at time t are overlap\\0,otherwise\end{matrix}\right\\\end{cases}.
% \end{align}

\noindent We then normalize $h_2(t)$ to $\bar{h}_2(t)$ by

\begin{equation}
        \bar{h}_{2}(t)=\frac{\sqrt{\int h_{3}^{2}(t)I_{23}(t) dt}}{\sqrt{\int h_{3}^{2}(t)I_{23}(t) dt}}h_{2}(t).\label{eq:17}
\end{equation}

\noindent We then normalize $h_1(t)$ to $\bar{h}_1(t)$ by

\begin{equation}
        \bar{h}_1(t)=\frac{\sqrt{\int\bar{h}_2^2(t)I_{12}(t) dt}}{\sqrt{\int h_1^2(t)I_{12}(t) dt}}h_1(t).\label{eq:18}
\end{equation}

\noindent We then construct the waveform $h(t)$ by

\begin{align}
        h(t)&=h'(t)-h''(t)\\
            &=h'(t)\biggl[1-\frac{1}{2}I_{12}(t)-\frac{1}{2}I_{23}(t)\biggr] , \notag \label{eq:21}
\end{align}

\noindent where $h'(t)$ and $h''(t)$ are defined as 

\begin{equation}
        h'(t)=\left[\bar{h}_{1}(t)+\bar{h}_{2}(t)+h_{3}(t)\right].\label{eq:19}
\end{equation}
\begin{align}
        h''(t)&=\frac{\bar{h}_{1}(t)+\bar{h}_{2}(t)}{2}I_{12}(t)+\frac{\bar{h}_{2}(t)+h_{3}(t)}{2}I_{23}(t)\\
        &=\frac{\bar{h}_{1}(t)+\bar{h}_{2}(t)+h_{3}(t)}{2}\big(I_{12}(t)+I_{23}(t)\big). \notag \label{eq:20}
\end{align}

\section{The performance of BNS waveform reconstruction\label{section:4}}

Three models, namely Model I, Model II, and Model III, were trained using the training dataset. Each model specializes in a specific stage: Model I focus on the waveform extraction of early inspiral stage, Model II on the waveform extraction of late inspiral stage, and Model III on the waveform extraction of near merger stage. In this section, we evaluate the performance of all three models using the test dataset.

To enhance the denoising model, we have upgraded the conventional UNet framework by incorporating Transformer and bridge elements to the feature output of the final Encoder Module. Additionally, we trained three Amplitude Regularity Models to facilitate a deeper analysis of the denoising model's output. To validate the utility of both the Transformer module and Amplitude Regularity Models, we also conducted performance tests with and without the Transformer.

We use overlap of the denoised waveform and the buried signal as the evaluation metric for waveform extraction. Suppose $a(t)$ and $b(t)$ are two timeseries with aligned phases, the overlap of $a(t)$ and $b(t)$ can be calculated as follows:

\begin{figure*}[htbp]
        \includegraphics[scale=1.1]{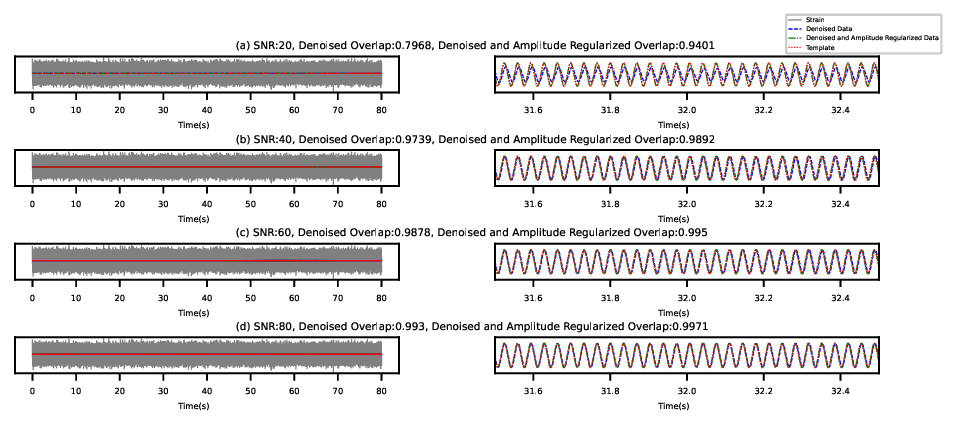}
        \caption{\label{fig:9}The comparisons between the denoising results and the signal on the early inspiral stage, at SNRs of 20, 40 60 80 sequentially.}
\end{figure*}

\begin{equation}
        overlap(a,b)=\frac{\int a(t)b(t) dt}{\sqrt{\int a^2(t)dt\int b^2(t)dt}}.\label{eq:22}
\end{equation}

Please note that Eq. (\ref{eq:22}) differs from its counterpart in previous works \cite{51,52}, and it has the capability to reflect both the phase alignment and amplitude variation of two signals. By computing the overlap metric, we can evaluate the model's denoising performance on the test datasets. The closer the metric is to 1, the more effectively the model can remove noise while retaining the BNS events from the original datasets. A lower metric indicates that the model may lose some original data during the denoising process.

For each waveform in the test dataset (30,000 waveforms), we initially normalized it to a randomly selected SNR value falling within the range of 20 to 80. Subsequently, we randomly trimmed the background noise and constructed the strain with combine of the signal and noise.

\begin{figure}[htbp]
        \includegraphics[scale=0.17]{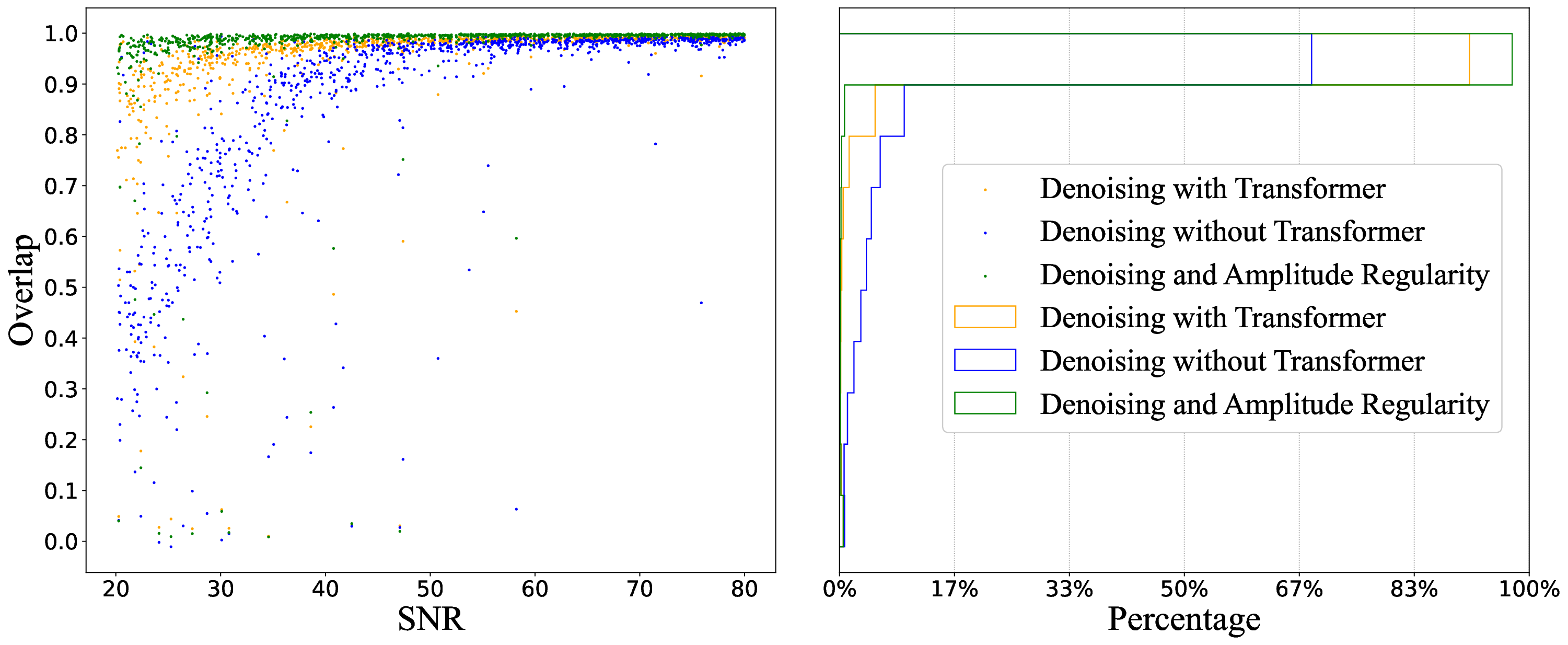}
        \caption{\label{fig:10}The overlap of the denoised output and the buried signal in test dataset on the early inspiral stage.}
\end{figure}

\subsection{Performance of waveform extraction on the early inspiral stage\label{Performance of waveform extraction on the early inspiral stage}}

We initially processed each combined strain of test data by passing it through a low-pass filter set at a cutoff frequency of 256 Hz. Following this, we resampled the strain data at a frequency of 512 Hz. Next, we randomly selected an 80-second segment of the resampled strain, ensuring that its endpoint preceded the merger time by 15 to 18 seconds. Subsequently, we fed this 80-second segment into Model I and obtained the output. 

The comparisons between the extracted results and the underlying signal of a randomly chosen resampled strain at various SNRs of 20, 40, 60, and 80 are presented in Fig. \ref{fig:9}. As depicted in Fig. \ref{fig:9}, at a SNR of 20, relying solely on the denoising model incorporating the Transformer leads to notable waveform distortion. Nevertheless, once the denoised signal is processed through the amplitude regularity model, the waveform is rectified, aligning more closely with the template. For SNRs exceeding 40, the substantial distortion in the denoised waveform diminishes significantly, regardless of whether the amplitude regularity model is employed. Macroscopically, there is no discernible difference in waveform distortion. 

\begin{figure*}[htbp]
        \includegraphics[scale=1.1]{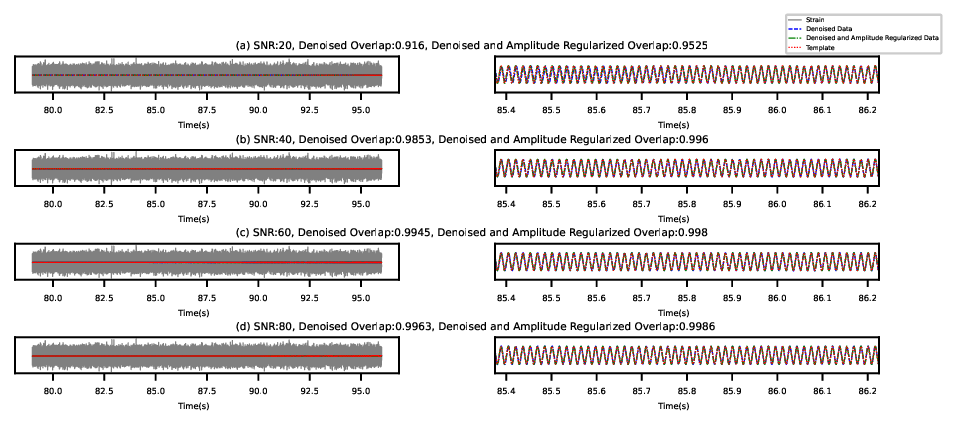}
        \caption{\label{fig:11}The comparisons between the denoising results and the signal on the late inspiral stage, at SNRs of 20, 40 60 80 sequentially.}
\end{figure*}

In Fig. \ref{fig:10}, we illustrate the overlap distribution between the extracted early inspiral waveform and the buried template. For comparison, we also depict the corresponding overlap distributions of the outputs from denoising models, comparing those with and without the inclusion of the Transformer module. In the absence of the Transformer module, only approximately 70\% of the samples exhibited an overlap exceeding 90\%. However, the introduction of the Transformer module significantly enhanced denoising performance, increasing the 90\% overlap percentage from 70\% to 90\%. Furthermore, when the Amplitude Regularity Model was incorporated, the waveform extraction performance rose further, the 90\% overlap percentage increased about 5\%. It is evident that the Transformer structure plays a pivotal role in significantly elevating the denoising capabilities of the model. Moreover, the Amplitude Regularity Model proves to be indispensable within our framework, exhibiting notable benefits, especially under low signal-to-noise ratio conditions.   

\begin{figure}[htbp]
        \includegraphics[scale=0.17]{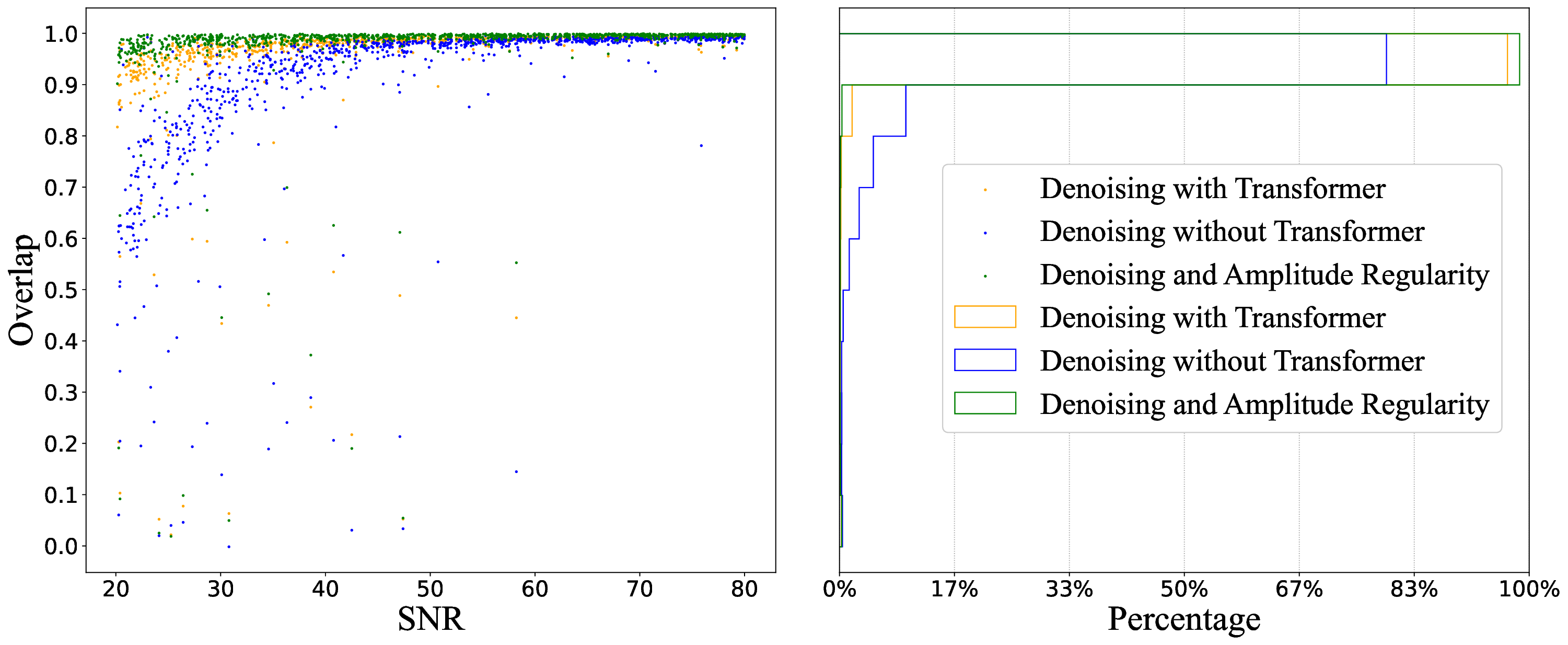}
        \caption{\label{fig:12}The overlap of the denoised output and the buried signal in test dataset on the late inspiral stage.}
\end{figure}

\subsection{Performance of waveform extraction on the late inspiral stage\label{Performance of waveform extraction on the late inspiral stage}}

In this subsection, akin to the preceding one, we conduct experiments aimed at evaluating the performance of Model II in waveform extraction. Initially, we applied a 1024 Hz low-pass filter to the combined strain, subsequently resampling it at a 2048 Hz frequency. Following this preprocessing step, we randomly cut the late inspiral part (the merger time being after the end time of the time slice about 1 s to 2 s) and proceeded to extract the buried signal.

Fig. \ref{fig:11} presents an example of the extracted output juxtaposed with the buried template. Evidently, in the denoising-only scenario at SNR 20, the overlap stands at 91.6\%. However, when we feed the denoised output into the Amplitude Regularity Model, the overlap is significantly enhanced to 95.25\%, marking an improvement of approximately 4\%. Akin to Fig. \ref{fig:10}, we have also analyzed the overlap distribution of Model II, as depicted in Fig. \ref{fig:12}. In the late inspiral case, both the transformer and amplitude regularity models prove crucial for effective waveform extraction.  

\begin{figure}[htbp]
        \includegraphics[scale=0.17]{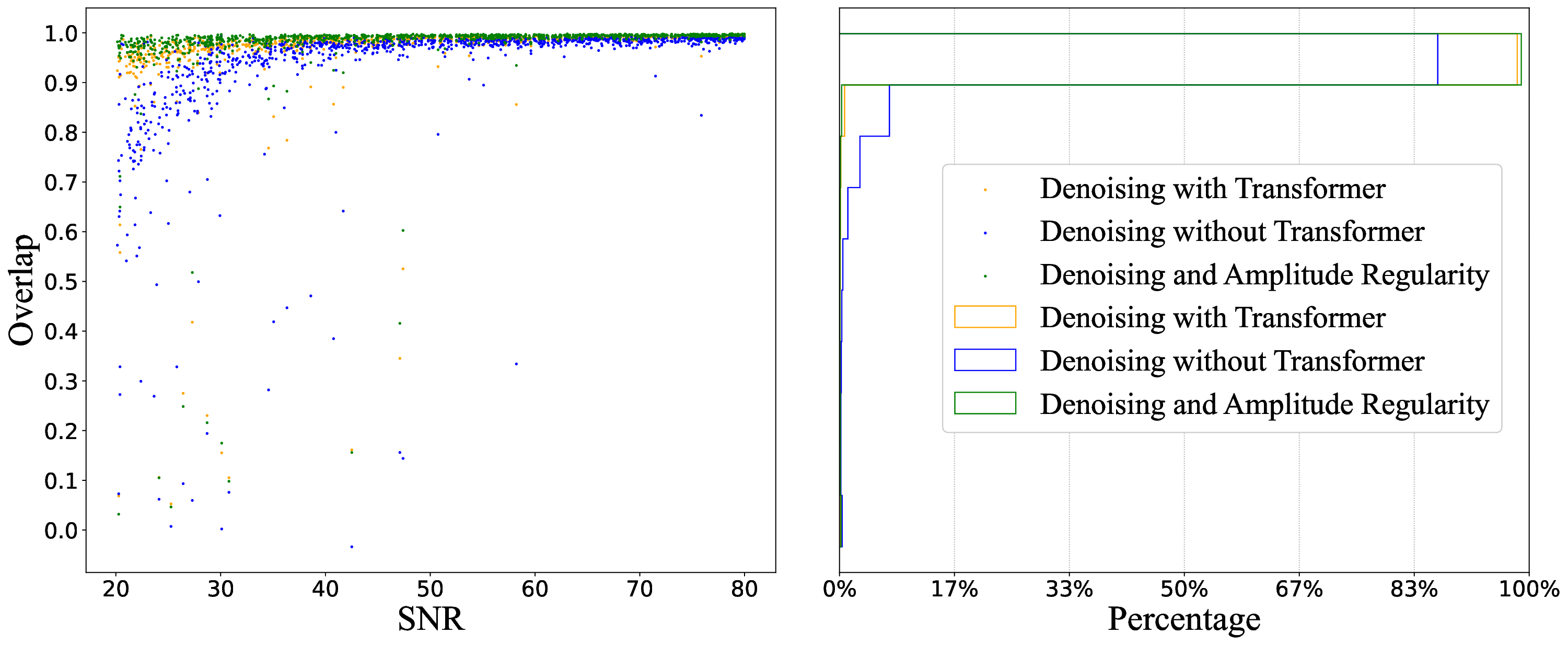}
        \caption{\label{fig:14}The overlap of the denoised output and the buried signal in test dataset near merger.}
\end{figure}

\begin{figure*}[htbp]
        \includegraphics[scale=1.1]{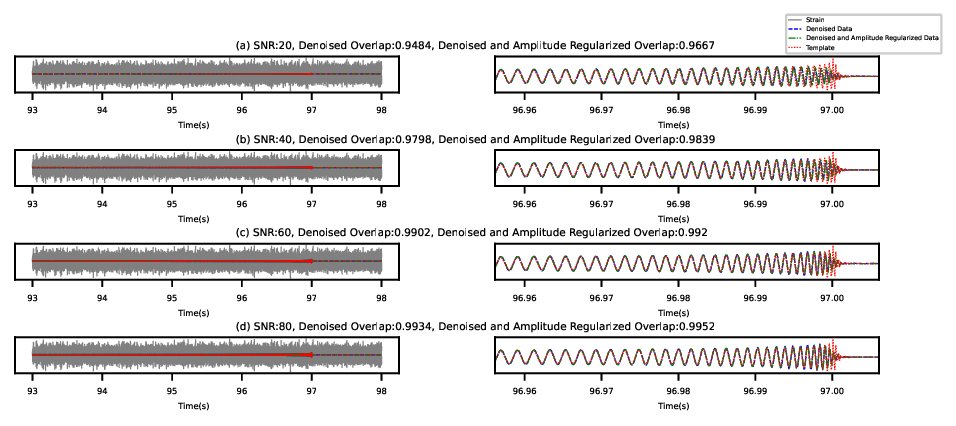}
        \caption{\label{fig:13}The comparisons between the denoising results and the signal ner merger, at SNRs of 20, 40 60 80 sequentially.}
\end{figure*}

\begin{figure*}[htbp]
        \includegraphics[scale=1.0]{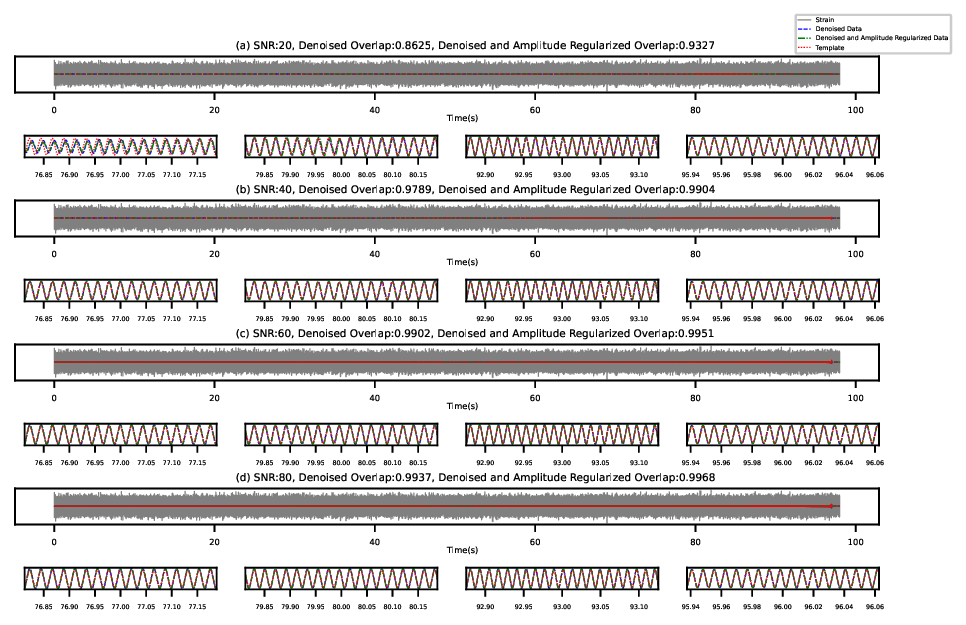}
        \caption{\label{fig:15}The comparisons between the denoising results and the signal after restructuring, at SNRs of 20, 40 60 80 sequentially.}
\end{figure*}

\subsection{Performance of waveform extraction near merger\label{Performance of waveform extraction near merger}}

We have previously analyzed the performance of waveform extraction during the early and late inspiral stages. In this subsection, we focus our attention on the waveform extraction performance specifically near the merger phase. To assess this, for each strain, we cut a 5-second segment containing merger phase, with the merger time occurring randomly between 3 to 4 seconds within the segment. We then extract the waveform from the selected segment to evaluate its performance.

As shown in Fig. \ref{fig:14} and Fig. \ref{fig:13}, the model exhibits performance similar to the preceding two inspiral periods before the merging time. However, during the peak interval of merging, the amplitude of the denoised signal fails to keep pace with the drastic changes in the template waveform, potentially resulting in premature attenuation of amplitude. This amplitude decay is more likely to occur under low SNR conditions.

\begin{figure}[htbp]
        \includegraphics[scale=0.17]{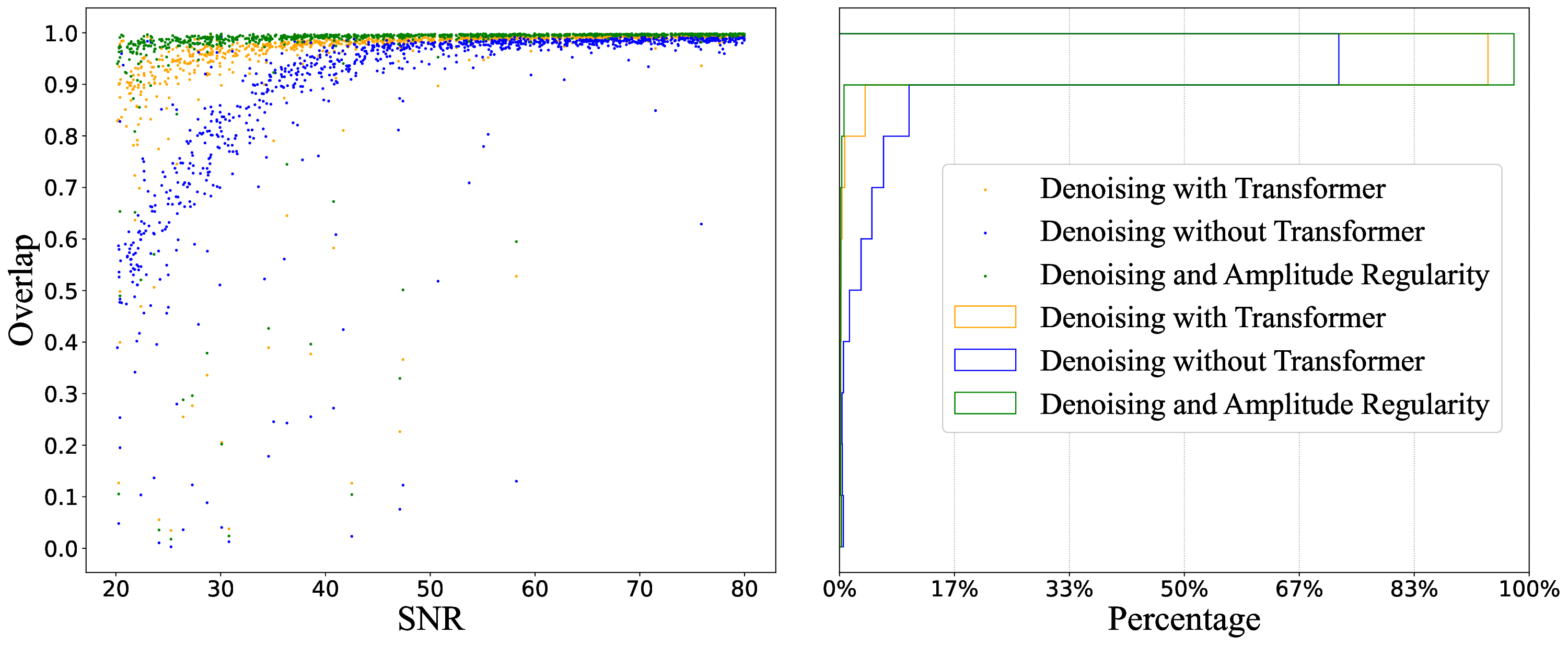}
        \caption{\label{fig:16}The overlap of the denoised output and the buried signal in test dataset for the whole BNS signal.}
\end{figure}

\subsection{Performance of reconstruction the whole 100 seconds length BNS signal\label{Performance of reconstruction the whole 100 seconds length BNS signal}}

At this stage, we reconstruct the data from the previous three stages into the original data and investigate the overall denoising effect. As depicted in Fig. \ref{fig:15} and Fig. \ref{fig:16}, this is a 98 second synthetic signal with overlaped positions between 77 to 80 seconds and 93 to 96 seconds. We also present waveform overlaped information for four overlaped edge positions. The results show that the designed waveform reconstruction method effectively integrate our segmented denoising results without sudden waveform fluctuations at the overlaped edge. It also adheres to the overall overlap observed in the previous three testing phases.

\section{Conclusion and Discussion\label{section:5}}

In conclusion, this study has presented a novel approach for BNS waveform extraction utilizing deep learning techniques. Through rigorous experimentation and analysis, we have demonstrated the effectiveness of our proposed method in addressing the challenges associated with processing GW data from ET. The proposed framework, inspired by the SPIIR method for matched filtering, combines denoising outputs of time-delayed strain to reconstruct the embedded BNS waveform. We have constructed three distinct denoising models, each tailored to specific phases of the BNS GW: early inspiral, later inspiral, and merger. We add Transformer encoder module to the Unet structured denoising model. We have conducted experiments to compare the denoising performance of the model both with and without the Transformer. The results demonstrate that the model incorporating the Transformer outperforms the one without it. We have also introduced the Amplitude Regularity Model, and experimental results indicate that this model can further enhance waveform extraction performance. We believe that our proposed method holds significant potential for early warning, searching, and localization of BNS GWs. 
Looking ahead, we anticipate that further refinements and extensions of our method could enhance its accuracy and reliability, ultimately contributing to a more comprehensive understanding of gravitational waves and their sources. We hope that our work will inspire future research in this exciting and rapidly developing field.

\nocite{*}

\bibliography{apssamp}% Produces the bibliography via BibTeX.

\end{document}